\documentclass{emulateapj}

\newcommand{\myemail}{endo@ioa.s.u-tokyo.ac.jp}
\newcommand{\smyr}{$M_{\mathrm{\sun}}\ \mathrm{yr}^{-1}$}
\newcommand{\Ldco}{$L\arcmin_{\mathrm{CO}}$}
\newcommand{\Mht}{$M_{\mathrm{H_2}}$}
\newcommand{\Sht}{$\Sigma_{\mathrm{H_2}}$}

\shorttitle{CO observation of the GRB 030329 host galaxy}
\shortauthors{Endo et al.}

\begin{document}


\title{A Revised Estimate of the CO ($J$ = 1--0) Emission from the Host Galaxy of GRB 030329 Using the Nobeyama Millimeter Array}


\author{A. Endo\altaffilmark{1,6}, K. Kohno\altaffilmark{1}, 
B. Hatsukade\altaffilmark{1},
K. Ohta\altaffilmark{2}, 
N. Kawai\altaffilmark{3,4},
Y. Sofue\altaffilmark{1}, 
K. Nakanishi\altaffilmark{5}, 
T. Tosaki\altaffilmark{5}, 
B. Vila-Vilaro\altaffilmark{6}, 
N. Kuno\altaffilmark{5}, 
T. Okuda\altaffilmark{1}, 
and K. Muraoka\altaffilmark{1} 
}


\altaffiltext{1}{Institute of Astronomy, the University of Tokyo, 2-21-1 Osawa, Mitaka, Tokyo 181-0015; \myemail}
\altaffiltext{2}{Department of Astronomy, Kyoto University, Kyoto 606-8502}
\altaffiltext{3}{Department of Physics, Tokyo Institute of Technology, 2-12-1 Ookayama, Meguro-ku, Tokyo 152-0033}
\altaffiltext{4}{Cosmic Radiation Laboratory, The Institute of Physical and Chemical Research (RIKEN), 2-1 Hirosawa, Wako, Saitama 351-0198}
\altaffiltext{5}{Nobeyama Radio Observatory, Minamimaki, Minamisaku, Nagano 384-1805}
\altaffiltext{6}{National Astronomical Observatory of Japan, 2-21-1 Osawa, Mitaka, Tokyo 181-8588}


\begin{abstract}
A sensitive observation of the CO ($J$ = 1--0) molecular line emission in the host galaxy of GRB 030329 ($ z =0.1685$) has been performed using the Nobeyama Millimeter Array in order to detect molecular gas and hidden star formation.
No sign of CO emission was detected, 
which invalidates our previous report on the presence of molecular gas. 
The 3$\sigma$ upperlimit on the CO line luminosity ($L\arcmin_{\mathrm{CO}}$) of the host galaxy is $6.9\times 10^8 \mathrm{\ K\  km\  s^{-1}\  pc^2} $. 
The lowerlimit of the host galaxy's metallicity is estimated to be $12+\log(\mathrm{O/H})\sim 
7.9$ ,
which yields a CO line luminosity to $\mathrm{H_2}$ conversion factor of $\alpha_{\mathrm{CO}} = 40 \ M_{\mathrm{\sun}}\ (\mathrm{K\ km\ s^{-1}\ pc^2})^{-1}$.
Assuming this $\alpha_{\mathrm{CO}}$ factor, the 3$\sigma$ upperlimit on the molecular gas mass of the host galaxy is $2.8 \times 10^{10}\  M_{\mathrm{\sun}}$. 
Based on the Schmidt law, the 3$\sigma$ upperlimit of the total star formation rate (SFR) of the host galaxy is estimated to be $38 \ M_{\mathrm{\sun}}\mathrm{\ yr^{-1}}$. These results independently confirm the inferences of previous observations in the optical, submillimeter, and X-ray band, which regard this host galaxy as a compact dwarf galaxy, and not a massive, aggressively star forming galaxy.
\end{abstract}


\keywords{galaxies: ISM ---  gamma rays: bursts --- gamma rays: individual (GRB 030329) ---  radio lines: galaxies}



\section{INTRODUCTION}

Long-duration gamma ray bursts (GRBs) provide a new and powerful means to detect distant galaxies and star formation. Due to their extremely energetic ($\sim$$10^{53} \ \mathrm{ergs}$) dust transparent gamma-ray emissions, GRBs can be detected from cosmological distances.
 It is now widely believed that GRBs accompany core-collapses of massive stars \citep[e.g.,][]{2003ApJ...591L..17S, 2003Natur.423..847H} and many GRBs have been found in star forming galaxies \citep[e.g.,][]{2003AJ....125..999B, 2003ApJ...588...99B, 2005NewA...11..103S}. 
This implies that the GRB formation rate (GFR) within a certain redshift bin is probably correlated with the star formation rate at that epoch \citep{1997ApJ...486L..71T, 1998MNRAS.294L..13W, 2000MNRAS.312L..35B}. If so, GRBs hold the potential to become extremely powerful tracers for distant star formation, extending the Madau diagram \citep{1996MNRAS.283.1388M} up to $z \gtrsim 20$ \citep{2000ApJ...536....1L}. However, there is no guarantee that the GFR is correlated only to the global SFR. The evolution of metallicity \citep{2006astro.ph..4113S,2006astro.ph..2444F} or the stellar initial mass function (IMF) in galaxies \citep{2002MNRAS.329..465R} could indeed influence the GFR. Of course, being able to track any of these parameters to cosmological distances will be very fruitful. Studying the basic properties of the GRB host galaxies is essential to precisely understand the type of evolution we are tracking by counting the GRBs.

GRB host galaxies often appear to be blue, sub-luminous, and low-metal dwarf galaxies in the optical/UV band, with a moderate SFR of $10^{-1} \sim 10^1\ M_{\sun}\ \mathrm{yr^{-1}}$
\citep{2003A&A...406L..63F, 2004A&A...425..913C, 2005NewA...11..103S}.
On the other hand,  \citet{2003ApJ...588...99B} have reported that some GRB host galaxies emit strongly in the submillimeter continuum, corresponding to SFRs higher by a few orders of magnitude compared to the SFRs measured in the optical/UV band, 
resembling active starforming galaxies enshrouded in heavy clouds of dust, 
commonly found at similar redshifts \citep{1999MNRAS.302..632B, 2001A&A...378....1F, 2001ApJ...556..562C}.
However, these GRB host galaxies were not detected by subsequent sensitive observations in the near/mid infrared window using the {\it Spitzer} space telescope  \citep{2006ApJ...642..636L}, which renders  conversion between the observed submillimeter fluxes and SFR suspect.
Clearly, confirmatory measurements in these wavelengths as well as more observations based on other independent methods are necessary for understanding this discrepancy.

Measuring the CO line luminosity (\Ldco ) of GRB host galaxies can possibly resolve this discrepancy. The correlation between a galaxy's \Ldco \ and its total molecular gas mass (\Mht ) is well known \citep{2005ARA&A..43..677S}. Once the \Mht \ and subsequently the molecular gas surface density (\Sht ) are known, its total SFR can be estimated using the global Schmidt law \citep{1998ApJ...498..541K}. Since millimeter waves penetrate through dust, the total SFR of a galaxy measured in this manner is
influenced neither by its dustiness nor its dust temperature. However, to date, no CO measurements of the \Mht \  of GRB host galaxies have been performed. This is mainly because GRBs are found at cosmological distances \citep[$z_{\mathrm{mean}} \sim 2.7$ for GRBs detected by {\it Swift};][]{2006astro.ph..2071J}, 
and their CO emission is too faint to be detected by the existing millimeter wave telescopes. 

Here, we present the results of an observation of the CO ($J$ = 1--0) molecular line emission in the host galaxy of GRB 030329 using the Nobeyama Millimeter Array (NMA). GRB 030329 is one of the GRBs that occurred closest to the Earth \citep[$z = 0.1685$; ][]{ 2003GCN..2020....1G, 2003GCN..2053....1C} and was the third closest one to us at the time of observation\footnote{http://www.mpe.mpg.de/~jcg/grbgen.html}. The optical image of the host galaxy appears to be a dwarf galaxy resembling the Small Magellanic Cloud (SMC) \citep[$M_{\mathrm{V}}\sim -16.5$, ][]{2003GCN..2243....1F} with a moderate extinction-corrected SFR of $\sim$0.5 \smyr \citep{2003ApJ...599..394M}. However, \citet[][K05, hereafter]{2005PASJ...57..147K} found a possible feature of CO ($J$ = 1--0) emission in the millimeter wave spectrum during an afterglow observation using the NMA, showing that the host galaxy possibly possesses an enormous amount of molecular gas ($M_{\mathrm{H_2}} > 10^9 \ M_{\sun}$). 
This amount is substantially more than what would be expected from the optical faintness of the host galaxy.
For example, the total molecluar gas mass of the SMC is only $\gtrsim$4 $\times 10^6 M_{\sun}$ \citep{2001PASJ...53L..45M}.
If this line actually exists then it would imply that a large amount of star formation in this galaxy is obscured by heavy dust clouds.
 This would certainly support the relationship between the GRBs and star formation. Therefore, the host galaxy of GRB 030329 was considered to be the best target to conduct the first ever deep CO ($J$ =1--0) observation.

Assuming a cosmology with $H_0 = 71 \ \mathrm{km \ s^{-1} \ Mpc^{-1}}$,  $\Omega_{\mathrm{M}}=0.27$, and $\Omega_{\Lambda}=0.73$, the luminosity distance of GRB 030329 is $d_{L}=802\mathrm{\ Mpc}$, and the angular distance is $d_{\mathrm{A}}=587\mathrm{\ Mpc}$ ($1\arcsec$ corresponds to 2.85\ kpc).

\section{OBSERVATIONS AND DATA ANALYSIS}
We conducted a set of sensitive observations of the CO ($J$ = 1--0) emission in the host galaxy of GRB 030329 using the NMA.
The observational log is presented in Table \ref{table1}.
The observation was performed in two periods that were separated by 4 months.
The first period was from 2004 December 9 to 14 and the second one was from 2005 April 22 to 29. The most compact configuration (D array; baseline lengths ranged from 13 m to 82 m) was used for the observations. 
The tracking frequency was set at 98.824 GHz or 98.905 GHz to observe the redshifted CO ($J$ = 1--0) line ($\sim$98.65 GHz) at the upper side band (USB).
The center frequency was shifted to ensure that the detected line feature is actually due to molecular emission and not due to any unexpected correlator  bandpass characteristics.
The lower side band (LSB) was centered at 86.825 GHz and 86.905 GHz, respectively. 
Each band was separated  by a $90\arcdeg $ phase switching of the reference signal. 
The Ultra-wide Band Correlator \citep{2000PASJ...52..393O} was configured to cover a bandwidth of 1024 MHz per side-band with a resolution of 8 MHz.

The radio sources B1741-038 and 3C 84 were observed every $\sim$20 minutes for amplitude and phase calibrations, and the passband shape of the system was determined based on the observations of the strong continuum sources -- 3C 84 or 3C 279. We also observed B1040+244 and B0953+254 several times during each observation run to verify the consistency of the amplitude calibration. The flux densities of B1040+244 and B0953+254 were determined several times during the observation runs; the flux density of B1040+244 ranged from 0.92 to 0.97\ Jy, and that of B0953+254 was 0.93\ Jy. In this study, we adopted a constant flux value of 1.0 Jy for both of these sources throughout the observation runs.

The raw visibilities were edited and calibrated using the NRO UVPROC-II package 
\citep{1997ASPC..125...50T}.
Images were produced using the NRAO AIPS task IMAGR. 
By applying different levels of smoothing to the observed data, with a native spectral resolution of 8 MHz $(24.3\ \mathrm{km\  s^{-1}})$, we produced series of intensity maps and spectrum with various spectral resolution.
The upperlimits on the CO flux, discussed in the subsequent sections, are based on the rms noise level measured over these intensity maps.

\begin{deluxetable}{lcccc}

\tabletypesize{\footnotesize}


\tablecaption{Nobeyama Millimeter Array Observations of GRB 030329}

\tablenum{1}

\tablehead{\colhead{UT date} & \colhead{Duration} & \colhead{$f_{\mathrm{USB}}$} & \colhead{$T_{\mathrm{sys}}$} & \colhead{Seeing} \\ 
\colhead{} & \colhead{(hr)} & \colhead{(GHz)} & \colhead{(K)} & \colhead{} \\
\colhead{(1)} & \colhead{(2)} & \colhead{(3)} & \colhead{(4)} & \colhead{(5)} 
} 

\startdata
2004 Dec 8 & 8.8  & 98.905  & 157  &  \\
2004 Dec 9 & 10.3  & 98.905  & 174  &  \\
2004 Dec 10 & 10.3  & 98.905  & 158  &  \\
2004 Dec 11 & 8.8  & 98.825  & 194  & Bad \\
2004 Dec 12 & 10.8  & 98.825  & 184  & Bad \\
2004 Dec 13 & 8.7  & 98.825  & 156  &  \\
 \\
2005 Apr 22 & 7.0  & 98.905  & 139  &  \\
2005 Apr 23 & 9.1  & 98.825  & 152  &  \\
2005 Apr 24 & 9.1  & 98.825  & 258  &  \\
2005 Apr 25 & 7.0  & 98.825  & 197  & Bad \\
2005 Apr 26 & 7.5  & 98.905  & 176  &  \\
2005 Apr 27 & 5.5  & 98.905  & 180  &  \\
2005 Apr 28 & 8.0  & 98.905  & 163  \\
\enddata

\tablecomments{(1) Observation date. (2) Observation Duration. The net on-source integration time is typically about half of this. (3) The center of the USB frequency. (4) System noise temperature in DSB. (5) Atmospheric stability. ``Bad'' implies that a large fraction of the visibility data ($\gtrsim$50 $\%$) was discarded due to significant phase noise.}



\label{table1}

\end{deluxetable}

\section{RESULTS}
Although we carefully examined the produced spectrum and intensity maps,  
no sign of CO emission from the host galaxy of GRB 030329 could be found, 
neither in the data that was newly acquired from our observation, nor in this data combined with those reported in K05.
As the significance of the possible emission feature reported earlier in K05
was merely 2  $\sigma$, we believe that in all likelihood it was purely noise.
In this section, we present the upperlimits placed on the total CO ($J$ = 1--0) flux of the host galaxy of GRB 030329.
\subsection{New Data}

Prior to adding our data to that of K05, we analyzed our data alone. This is because (1) our observation was conducted more than a year after the burst; thus, the possibility of contamination from the emission of the burst is excluded, although the continuum component of the emission was properly subtracted in K05 
as well.
 (2) Further, the band pass characteristics of the instruments used for observation drift over the years and the two sets of data can be regarded as being obtained from two independent instruments, at least to some extent. Therefore, the separate analysis of the data  enables a cross-check for any instrumental errors.

Because the velocity width ($\Delta v$) of the CO line could not be measured, we assumed two different velocity widths to estimate the upperlimit of the velocity integrated CO line flux ($I_{\mathrm{CO}} = \int S_{\mathrm{CO}}\ dv$, where $S_{\mathrm{CO}}$ denotes the CO flux density per beam). First, we adopted $\Delta v=220\ \mathrm{km\ s^{-1}}$, which was the velocity width of the emission feature reported in K05. In this case, a 3$\sigma$ upperlimit of $3\cdot \delta I_{\mathrm{CO}} = 0.96\   \mathrm{Jy\   km \  s^{-1}}$ can be set on the host galaxy. This is less than half of the upperlimit attained in K05.

Since the CO ($J$ = 1--0) emission line was not detected, the host galaxy of GRB 030329 is most likely to be a dwarf galaxy as observed in the optical band \citep{ 2003GCN..2243....1F, 2003ApJ...599..394M, 2005A&A...444..711G, 2005NewA...11..103S}. For this reason, we next adopted $\Delta v = 95 \mathrm{\ km\ s^{-1}}$ as a more realistic estimation of the velocity width of the host galaxy. This is the average of the 115 dwarf galaxies selected by \citet{2005ApJ...625..763L} with rotation velocities of $v_{\mathrm{rot}} = \ 67  \mathrm{km \  s^{-1}}$, multiplied by $\sin 45\arcdeg$ assuming an inclination angle of $45\arcdeg$. In this case, we derive $3 \cdot \delta I_{\mathrm{CO}} = 0.60\  \mathrm{Jy \  km \  s^{-1}}$.

\subsection{Combined Data With \citet{2005PASJ...57..147K}}
\begin{figure}
\epsscale{1}
\plotone{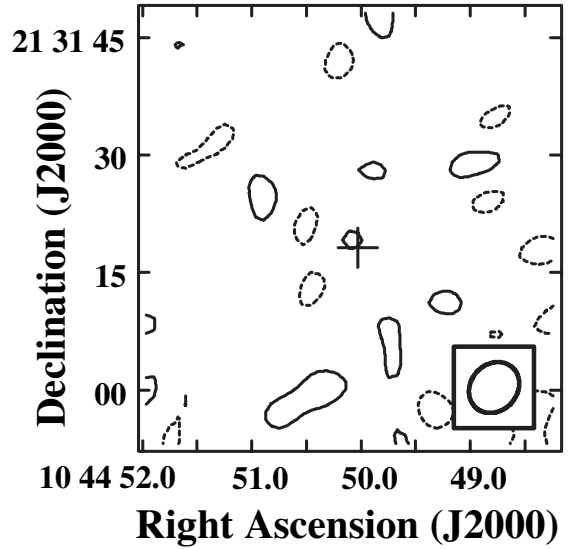}
\caption{Intensity map of the direction of GRB 030329 obtained after combining our data with that from \citet{2005PASJ...57..147K}. 
The intensity is integrated over a velocity width of $220\mathrm{\ km\ s^{-1}}$, to allow direct comparison to Figure 3 in \citet{2005PASJ...57..147K}.
The synthesized beam, shown in the right bottom corner, is 6$\farcs$95 $\times$ 5$\farcs$94 (position angle = $-43 \arcdeg$). The contour interval is 0.46$\  \mathrm{Jy\ beam^{-1}\  km \  s^{-1}}\ (1.5\sigma)$. 
Solid contours indicate positive fluxes, and broken contours indicate negative fluxes.
The cross hair at the center coincides with the position of the GRB 030329 optical afterglow ($\mathrm{\alpha_{J2000} = 10^h44^m50.03}$,  $\mathrm{\delta_{J2000} = +21\arcdeg 31 \arcmin18 \farcs 15})$.
The emission feature at the center has almost disappeared ($\la 1.5\sigma$), 
and its significance could not be improved by increasing spectral resolution.
\label{fig1}}
\end{figure}

\begin{figure}
\epsscale{1}
\plotone{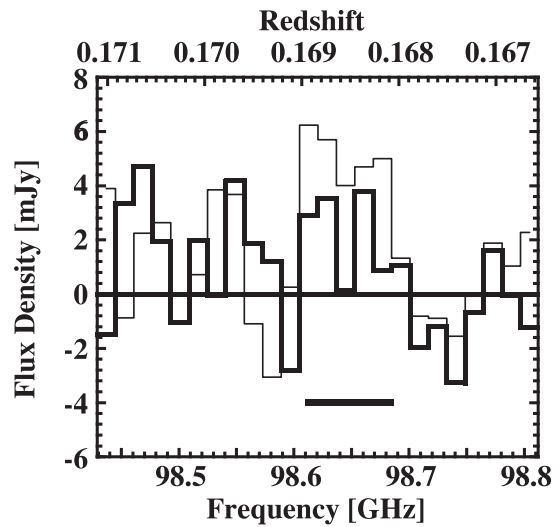}
\caption{Spectrum observed in the 98  GHz band toward GRB 030329. 
The thin lines represent the spectrum presented in \citet{2005PASJ...57..147K}, and the thick line shows the spectrum produced by combining our data with those of \citet{2005PASJ...57..147K}. 
The spectrum is smoothed to a resolution of 16 MHz or $\mathrm{48.6\   km\   s^{-1}}$. The horizontal bar indicates the velocity range where a possible emission feature is visible in the old spectrum centered in the redshifted CO ($J$ = 1--0) emission frequency (98.65 GHz) of GRB 030329 ($z = 0.1685$). The rms noise level has improved from 3.8 $\mathrm{mJy\ beam^{-1}}$ to 2.0 $\mathrm{mJy\ beam^{-1}}$, 
and the emission feature has lost its significance.
The significance could not be improved by changing the smoothing factor.
\label{fig2}}
\end{figure}

Since no CO line emission features were detected in our data, we combined our data with the K05 data in order to obtain the lowest possible upperlimit on the total flux.
Figure \ref{fig1} shows the intensity map of the direction of GRB 030329 integrated over $\Delta v = 220 \ \mathrm{km\ s^{-1}}$.
Figure \ref{fig2} shows the spectrum in the 98 GHz band toward GRB 030329.
The significance of the emission feature reported in K05 has 
been 
reduced to near the rms noise level ($\sim$1.5 $\sigma$).
We again place an upperlimit on the velocity integrated flux density of the CO emission from the host galaxy using the two abovementioned candidates for estimating the line velocity width. Assuming $\Delta v = 220 \  \mathrm{km\   s^{-1}}$, the 3$\sigma$ upperlimit of $I_{\mathrm{CO}}$ is $3\cdot \delta I_{\mathrm{CO}} = 0.89 \  \mathrm{Jy \  km\  s^{-1}}$. Likewise, $\Delta v = 95 \  \mathrm{km \  s^{-1}}$ yields an upperlimit of $3\cdot\delta I_{\mathrm{CO}} = 0.51 \  \mathrm{Jy\   km \  s^{-1}}$. We will adopt $3\cdot \delta I = 0.51 \  \mathrm{Jy\   km \  s^{-1}}$ in the subsequent sections, assuming that the host galaxy is an average dwarf galaxy.

\section{DISCUSSION}
\subsection{Molecular Gas Mass}

\begin{deluxetable*}{ccccc}


\tabletypesize{\scriptsize}


\tablecaption{Influence of Different Values of Host Galaxy Metallicity upon the Results}

\tablenum{2}

\tablehead{  \colhead{$12+\log(\mathrm{O/H})$} & \colhead{$\alpha_{\mathrm{CO}}$} & \colhead{$3\cdot \delta M_{\mathrm{H_2}}$}& $3\cdot \delta \Sigma_{\mathrm{H_2}}$ &$3\cdot \delta \mathrm{SFR}$
\\ 
 \colhead{} & \colhead{($M_{\mathrm{\odot}}\ (\mathrm{K\ km\ s^{-1}\ pc^2})^{-1}$)} 
& \colhead{($10^{10}\ M_{\mathrm{\sun}}$)}
& \colhead{$(M_{\mathrm{\sun}}\ \mathrm{pc^{-2}})$}
& \colhead{$M_{\sun}\ \mathrm{yr^{-1}}$}
\\
\colhead{(1)} & \colhead{(2)} & \colhead{(3)} & \colhead{(4)} & \colhead{(5)} }

\startdata
7.9&    40 &    $2.8\times 10^{10}$ &    72 &    38 \\
8.3&    16 &    $1.1\times 10^{10}$&    29 &    11 \\
8.6&    7.9 &    $5.5\times 10^{9}$ &    14 &    4.0 \\
\enddata


\tablecomments{Col. (1): Possible values of the host galaxy's metallicity; 
Col. (2): The CO line luminosity to $\mathrm{H_2}$  molecular gas mass conversion factor, calculated using  equation (\ref{metal}); 
Col. (3): The 3$\sigma$ upperlimit of the total molecular gas mass, calculated using equation (\ref{gasmass}); 
Col. (4): The 3$\sigma$ upperlimit of the molecular gas mass column density; 
Col. (5):  The 3$\sigma$ upperlimit of the total star formation rate of the host galaxy, calculated using equation (\ref{SFR}).}

\label{tab_metal}
\end{deluxetable*}

\begin{deluxetable*}{rccccc}


\tabletypesize{\scriptsize}


\tablecaption{Constraint on the Molecular Gas Mass of the Host Galaxy of GRB 030329}

\tablenum{3}

\tablehead{  \colhead{}& \colhead{$\delta S_{\mathrm{CO}}$} & \colhead{$3\cdot \delta L\arcmin_{\mathrm{CO}}$} & \colhead{$3\cdot \delta M_{\mathrm{H_2}}$}& $3\cdot \delta \Sigma_{\mathrm{H_2}}$ & Ref.
\\ 
\colhead{} & \colhead{($\mathrm{mJy \ beam^{-1}}$)} & \colhead{$(10^8\ \mathrm{K\  km\  s^{-1}\  \mathrm{pc^2}})$} 
& \colhead{($10^{10}\ M_{\mathrm{\sun}}$)}
& \colhead{$(M_{\mathrm{\sun}}\ \mathrm{pc^{-2}})$}
& \colhead{}
\\
\colhead{} &\colhead{(1)} & \colhead{(2)} & \colhead{(3)} & \colhead{(4)} & \colhead{(5)} }

\startdata
$\mathrm{1^{st}}$ run & 4.0 & $15 $ & $6.1$
& $1.6 \times 10^2$& 1\\
$\mathrm{2^{nd}}$ run & 2.1 & $8.1 $ & $3.2 $
& 85 & this study\\
total & 1.8 & $6.9 $ &$2.8 $
& 72 & this study\\
\hline
from OH absorption &  &  &  
& 112& 2\\
\enddata


\tablecomments{Col. (1): The rms noise level (1$\sigma$) of the flux density per beam. The spectral resolution is 32 MHz, which corresponds to a velocity width of $97 \mathrm{\ km \  s^{-1}}$; col. (2): The 3$\sigma$ upperlimit of the CO($J$ = 1--0) line luminosity, assuming a line velocity width of $\Delta v = 95 \  \mathrm{km \  s^{-1}}$; col. (3): The 3$\sigma$ upperlimit of the total molecular gas mass, assuming $\alpha_{\mathrm{CO}}=40 \ M_{\mathrm{\sun}}\ (\mathrm{K\ km\ s^{-1}\ pc^2})^{-1}$; col. (4): The 3$\sigma$ upperlimit of the molecular gas mass column density. The values derived from the CO line observation are averaged within the synthesized beam. The value derived from OH absorption is the density within the line of sight toward the GRB, assuming a galactic $\mathrm{OH}$/$\mathrm{H_2}$ abundance value; 
col. (5): References.\label{tab_sum}}

\tablerefs{(1) \citealt{2005PASJ...57..147K}; (2) \citealt{2005ApJ...622..986T}}
\end{deluxetable*}

The total molecular gas mass ($M_{\mathrm{H_2}}$) of a galaxy is proportional to its CO line luminosity (\Ldco ); $M_{\mathrm{H_2}} = \alpha_{\mathrm{CO}}L\arcmin_{\mathrm{CO}}$, where $\alpha_{\mathrm{CO}}$ is the CO line luminosity to $\mathrm{H_2}$   molecular gas mass conversion factor ($\alpha_{\mathrm{CO}}$ factor). Here, we make an attempt to set an upperlimit on the molecular gas mass of the host galaxy of GRB 030329.

The 3$\sigma$ upperlimit of the $L\arcmin_{\mathrm{CO}}$ can be calculated from the obtained $\delta I_{\mathrm{CO}}$, using the following formula \citep{2005ARA&A..43..677S}:
\begin{eqnarray}
3 \cdot \delta L\arcmin_{\mathrm{CO}} &=& \Biggl(\frac{c^2}{2k} \Biggr)\cdot (3\cdot \delta I_{\mathrm{CO}}) \cdot d_L^2 \cdot \nu_{\mathrm{rest}}^{-2} (1+z)^{-1} \nonumber \\
&=& 6.9 \times 10^{8} \  \Biggl(\frac{3\cdot \delta I_{\mathrm{CO}}}{0.51 \  \mathrm{Jy\  km\  s^{-1}}}\Biggr)
\Biggl(\frac{\nu_{\mathrm{rest}}}{115\ \mathrm{GHz}} \Biggr)^{-1}
\nonumber \\
\times&
\Biggl(&\frac{d_{\mathrm{L}}}{802\  \mathrm{Mpc}}\Biggr)^2  \Biggl(\frac{1+z}{1.1685} \Biggr)^{-1} 
\mathrm{K\  km\  s^{-1} pc^2}.
\end{eqnarray}

The uncertainty of the $\alpha_{\mathrm{CO}}$ factor is one of the largest error sources of the CO-measured molecular gas mass. There is a large variation of about two orders of magnitude in the global $\alpha_{\mathrm{CO}}$ factor among galaxies. The dwarf galaxies, which have poor metallicity, tend to have large $\alpha_{\mathrm{CO}}$ factors \citep[e.g., $\alpha_{\mathrm{CO}} = 40 \ M_{\mathrm{\sun}}\ (\mathrm{K\ km\ s^{-1}\ pc^2})^{-1}$ for SMC; ][]{2001PASJ...53L..45M} as compared to the Galactic value \citep[$\alpha_{\mathrm{CO}} = 2.9 \ M_{\mathrm{\sun}}\ (\mathrm{K\ km\ s^{-1}\ pc^2)^{-1}}$; ][]{2001ApJ...547..792D}. On the other hand, the nearby metal-rich ultra-luminous infrared galaxies (ULIRGs) typically have an $\alpha_{\mathrm{CO}}$ factor of $\lesssim$1 $M_{\mathrm{\sun}}\ (\mathrm{K\ km\ s^{-1}\ pc^2})^{-1}$ \citep{1998ApJ...507..615D}. 

Although the $\alpha_{\mathrm{CO}}$ factor of the GRB 030329 host galaxy is unknown, we can estimate it by using the following correlation between the metallicity and the $\alpha_{\mathrm{CO}}$ factor, which holds for the nearby dwarf and spiral galaxies \citep{1996PASJ...48..275A}.

\begin{equation}
\log \alpha_{\mathrm{CO}} = -1.0 \ \Biggl\{ 12+\log \Biggl( \frac{\mathrm{O}}{\mathrm{H}} \Biggr) \Biggr\} + 9.5.
\label{metal}
\end{equation}

\citet{2005NewA...11..103S} conducted a series of optical photometry and spectroscopy of the host galaxy of GRB 030329, and successfully measured the $R_{23} = ([\mathrm{OII}]\ \lambda 3727+ [\mathrm{OIII}]\ \lambda\lambda 4959, 5007)/\mathrm{H\beta}$ emission line ratio \citep[see also][]{2005A&A...444..711G}. This allowed them to perform the standard $R_{23}$ diagnostic \citep{2002ApJS..142...35K} and  estimate the metallicity of the host galaxy. However, the $R_{23}$ ratio has a maximum value at $12 + \log(\mathrm{O}/\mathrm{H})\sim 8.5$, and the result is two-valued. 
According to \citet{2005NewA...11..103S}, the lower and upper branch values are $12 + \log(\mathrm{O}/\mathrm{H})= 7.9$ and 8.6, respectively.
Though it is customary to break this degeneracy using other line ratios such as $[\mathrm{NII}]\ \lambda 6584/[\mathrm{OII}]\ \lambda 3727$, they were not able to do so because unfortunately none of these pairs came above their detection limit. Instead, they used the metallicity-luminosity relation presented by \citet{2003A&A...401..141L} and the  host galaxy's $M_B$ magnitude 
 \citep[$M_B\sim-16.5$;][]{2005A&A...444..711G}. This yields a host galaxy metallicity closer to the lower branch value, i.e., ; $12+\log(\mathrm{O/H})\sim 8.1$. Therefore, \citet{2005NewA...11..103S} state that the lower branch value is more plausible. 
Recently, a stronger upperlimit on the $[\mathrm{NII}]\ \lambda 6586$ has been reported by \citet{2006astro.ph.11772T}, which also supports the lower branch value.

However, a recent update of the metallicity-luminosity relation using $\sim$53,000  star-forming galaxies in the Sloan Digital Sky Survey yields a metallicity of $\sim$0.2 dex higher than the relation reported by \citet{2003A&A...401..141L} for a given $M_B$ magnitude \citep{2004ApJ...613..898T}. 
Adopting this relation, the host galaxy metallicity is estimated to be $12+\log(\mathrm{O/H})=8.3\pm 0.16$, which falls equally close to the two $R_{23}$ solutions. 
Moreover, it should be noted that the $R_{23}$ method rapidly loses sensitivity near its local maximum at $12+\log(\mathrm{O/H})\sim 8.5$. 
For example, applying $R_{\mathrm{23}}=6.1$ and $[\mathrm{OIII}]\lambda\lambda 4950, 5008 / [\mathrm{OII}]\lambda 3727 = 2.8 $ \citep{2005NewA...11..103S} to the calibration diagram presented by \citet[][Figure 8]{1999ApJ...514..544K} yields two closely neighboring solutions at around
$12+\log(\mathrm{O/H})\sim 8.3$-$8.6$ with errors of $\pm 0.25$ dex each.
Therefore, we must say that there is quite a large ambiguity in the host galaxy's metallicity, ranging from $12+\log(\mathrm{O/H}) \sim 7.9$ to $\sim$8.6. This difference has a considerable effect on the upperlimits of the molecular gas mass and SFR of the host galaxy, as summarized in Table \ref{tab_metal}. We will adopt $12+\log(\mathrm{O/H})=7.9$ in the subsequent discussion, which yields the most conservative upperlimits for both of these values. Substituting this in equation (\ref{metal}) yields $\alpha_{\mathrm{CO}}=40\ M_{\mathrm{\sun}}\ (\mathrm{K\ km\ s^{-1}\ pc^2})^{-1}$.

Applying the above-mentioned values of $3 \cdot \delta L\arcmin_{\mathrm{CO}}$ and $\alpha_{\mathrm{CO}}$, the 3$\sigma$ upperlimit of the total molecular gas mass of the host galaxy of GRB 030329 can be derived as follows:

\begin{eqnarray}
3\cdot\delta M_{\mathrm{H_2}} &=& \alpha_{\mathrm{CO}} \cdot (3\cdot \delta L\arcmin_{\mathrm{CO}}) \nonumber \\
&=& 2.8 \times 10^{10} \cdot \Biggl(\frac{\alpha_{\mathrm{CO}}}{40 \ M_{\sun} \ (\mathrm{K\  km\  s^{-1}\  pc^2)^{-1}}}\Biggr) \nonumber \\
&\ &\times \Biggl(\frac{3\cdot \delta L\arcmin_{\mathrm{CO}}}{6.9 \times 10^8\  \mathrm{K\  km\  s^{-1}\  pc^2}}\Biggr)\ M_{\sun}.
\label{gasmass}
\end{eqnarray}
Assuming a Gaussian profile, the synthesized beam (6$\farcs$95 $\times$ 5$\farcs$94, FWHM) corresponds to a surface area of $S=3.8\times 10^2\ \mathrm{kpc^2}$ at $z = 0.1685$. After adopting this value, the 3$\sigma$ upperlimit of the average molecular gas mass surface density ($\Sigma_{\mathrm{H_2}}$) within the beam can be calculated as $72 \ M_{\sun} \ \mathrm{pc^{-2}}$.

\citet{2005ApJ...622..986T} measured the OH main line opacity of the host galaxy of GRB 030329 and 
set an upperlimit of $3\cdot \delta \Sigma_{\mathrm{H_2}}=112 \ M_{\sun}\ \mathrm{pc^{-2}}$ using the relation $N_{\mathrm{OH}}/N_{\mathrm{H_2}}\approx 1\times 10^{-7}$. 
However, so far, this relation holds only for Galactic dark clouds \citep{1999ASPC..156..188L} and the four intermediate-redshift molecular absorption systems \citep{2002A&A...381L..73K}. 
This ratio may be lower in a galaxy with a low metallicity, thereby resulting in an even higher upperlimit on $\Sigma_{\mathrm{H_2}}$. 
Nevertheless,  the upperlimit we have set on the $\Sigma_{\mathrm{H_2}}$ of the host galaxy of GRB 030329 is the lowest ever to our knowledge.

\subsection{Star Formation Rate}

We can place a constraint on the total SFR of the host galaxy of GRB 030329, using the upperlimit of the $\Sigma_{\mathrm{H_2}}$ derived above.
Using the global Schmidt law \citep{1998ApJ...498..541K}, which holds for spiral and dwarf galaxies on a large scale \citep[e.g., ][]{2005PASJ...57..733K, 2005ApJ...625..763L}, the 3$\sigma$ upperlimit of the total SFR of the host galaxy can be deduced as follows:
\begin{eqnarray}
3\cdot \delta \mathrm{SFR} &=& 
38 
\times \Biggl(\frac{3\cdot \delta\Sigma_{\mathrm{gas}}}{
72
 \ M_{\sun}\ \mathrm{pc^{-2}}}\Biggr)^{1.4} \nonumber\\
&\ & \times \Biggl(         
\frac{S}{3.8\  \cdot 10^2 \mathrm{\ kpc^2}}
\Biggr)  
  \ M_{\sun}\ \mathrm{yr^{-1}}.
\label{SFR}
\end{eqnarray}
Note that adopting a smaller $\alpha_{\mathrm{CO}}$ factor of 
7.9
$M_{\sun}\ (\mathrm{K\ km\ s^{-1}\ pc^2)^{-1}}$ leads to a considerably more rigid upperlimit of $3\cdot \delta\mathrm{SFR}= 4.0 \ M_{\sun}\ \mathrm{yr^{-1}}$.
The influence of different possible values of the host galaxy's metallicity is presented in Table \ref{tab_metal}, where a summary of results from several studies are presented in Table \ref{tab_sum}.

\begin{deluxetable}{lccc}


\tabletypesize{\scriptsize}


\tablecaption{SFR of the Host Galaxy of GRB 030329}

\tablenum{4}

\tablehead{\colhead{Observational Method} & \colhead{SFR} & \colhead{Ref.} \\ 
\colhead{} & \colhead{($M_{\sun}\ \mathrm{yr^{-1}}$)} & \colhead{}  \\
\colhead{(1)} & \colhead{(2)} & \colhead{(3)}  }

\startdata
CO ($J$ = 1--0) & $<$38 \tablenotemark{a} & This study \\ 
$[\mathrm{OII}]$ 3727 and Balmer lines & $\sim$0.6\tablenotemark{b} &  1 \\
Photometric SED & $\sim$0.54 \tablenotemark{b} &  1 \\
Submillimeter continuum & $<$200 \tablenotemark{c} & 2 \\
X-ray luminosity & $\lesssim$200 $\pm 80$ & 3\\
\enddata


\tablenotetext{a}{Adopting $\alpha_{\mathrm{CO}} = 8.6 \ M_{\mathrm{\sun}}\ (\mathrm{K\ km\ s^{-1}\ pc^2)^{-1}}$ leads to a more rigid upperlimit of $<$4.0$\ M_{\sun}\ \mathrm{yr^{-1}}$.}
\tablenotetext{b}{Assuming the SMC extinction law ($A_{V}\sim$ 0.6).}
\tablenotetext{c}{We calculated the SFR based on the reported 3$\sigma$ upperlimit of the submillimeter flux density using the relation between the submillimeter flux density and the star formation rate \citep{1999ApJ...513L..13C}.}

\tablecomments{Col. (1): The observed emission feature; 
col. (2): The measured star SFR; 
col. (3): The instrument used for observation; 
col. (4): References.}

\tablerefs{
(1)\citealt{2005A&A...444..711G}; 
(2)\citealt{2005A&A...439..981S}; 
(3)\citealt{2004A&A...425L..33W}.
}

\label{table3}
\end{deluxetable}

Recently, there have been reports on the submillimeter \citep{2005A&A...439..981S} and X-ray \citep{2004A&A...425L..33W} observations of the host galaxy of GRB 030329
as well as optical observations corrected for extinction \citep{2005A&A...444..711G}.
Table \ref{table3}
lists the SFRs of the host galaxy measured with various star formation tracers.
The upperlimits derived from non-detections are included. 
Our upperlimit, which is derived from the molecular gas mass measured by the CO line luminosity, is the most rigid among the 
observational methods that are intrinsically free of dust extinction, even when the uncertainty of the 
$\alpha_{\mathrm{CO}}$ factor is considered.
This result confirms the moderate SFR ($\ll 100 \ M_{\sun}\ \mathrm{yr^{-1}}$) of the host galaxy.

\subsection{What is the True SFR of GRB Host Galaxies?}

\begin{figure*}
\epsscale{1}
\plotone{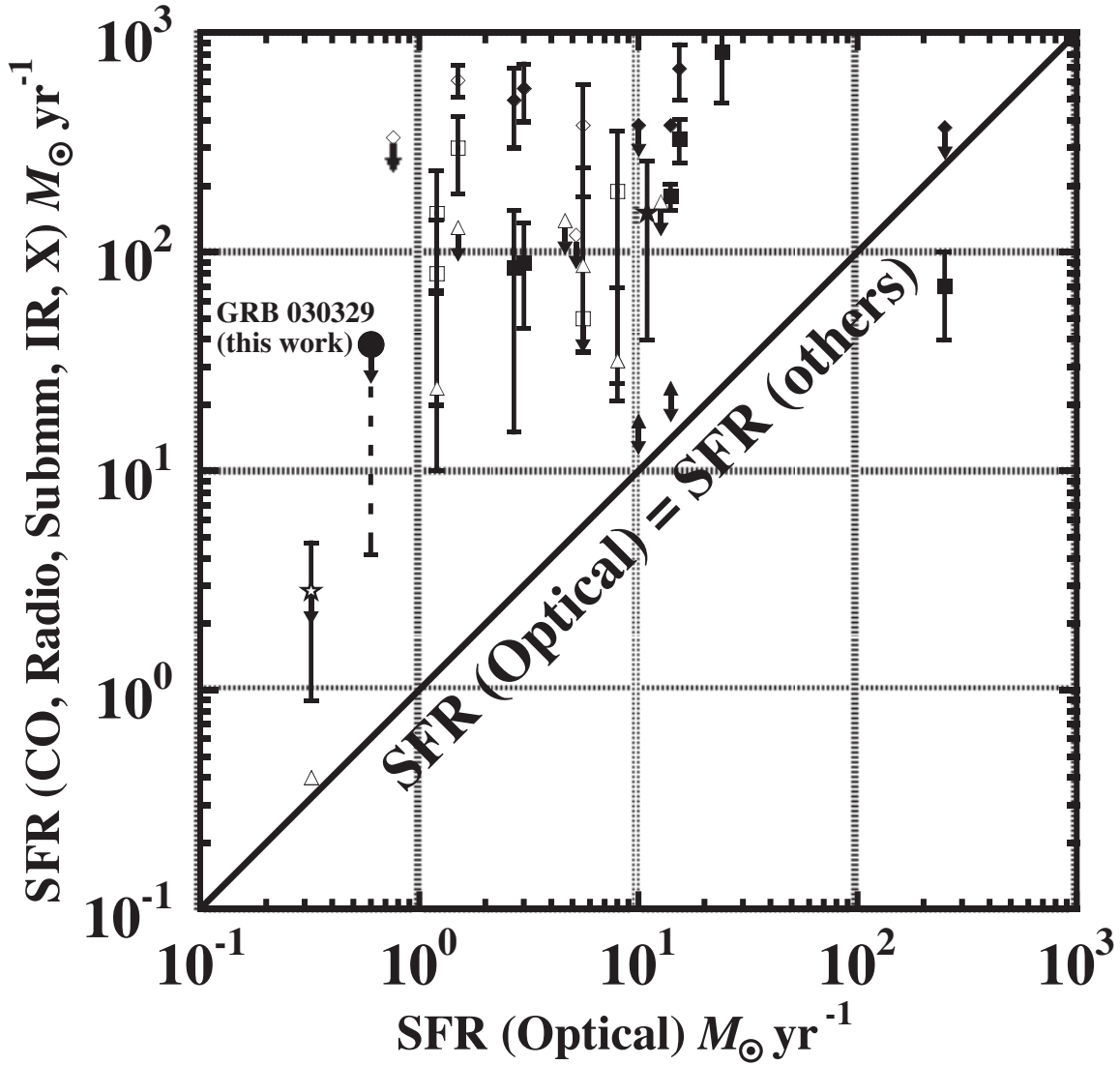}
\caption{\footnotesize{A comparison between the SFR of GRB host galaxies measured by optical methods and other methods that are free of dust extinction. References are listed in Table \ref{tab_studies}. The abscissa is the SFR measured by optical methods, namely, 
$\mathrm{[OII]}\ \lambda 3727$ and Balmer line luminosities and UV continuum luminosity. The ordinate is the SFR measured by various tracers in wavelengths other than those belonging to the optical/UV band, namely, CO line luminosity (circles; this study), radio (squares), submillimeter continuum luminosity (tilted squares), mid/far infrared continuum luminosity (triangles), and X-ray luminosity (stars).
Solid symbols have optical SFRs corrected for dust extinction, and the open ones are not corrected for extinction
in the host galaxy. The symbols with arrows are upperlimits for the host galaxies without any significant detection in the extinction free wavelength. The solid line indicates a one-to-one correspondence between the two SFRs.
The broken line indicates the uncertainty of the upperlimit measured by the CO line luminosity, 
which originates due to the uncertainty of metallicity and the $\alpha_{\mathrm{CO}}$ factor.
It is observed that SFRs measured using optical/UV wavelengths are significantly smaller compared to other methods, even after correction for dust extinction.}
\label{fig_SFR}}
\end{figure*}

Many observations of GRB  host galaxies in the optical/UV range infer that GRB host galaxies are blue, sub-luminous  dwarf galaxies with only a moderate SFR. However, SFR measurements in the optical/UV band can underestimate the true SFR if severe extinction by dust is present. Recently, there have been more such studies that make an attempt to correct for dust extinction, mostly using optical SED fits \citep[e.g.,][]{2001A&A...372..438S}. Such corrections are useful up to an extent, but  lose efficiency when the clouds of dust are so thick that the optical emission is completely obscured from view.

Recently, many groups have begun to measure the SFRs of GRB host galaxies using various methods that are immune to dust extinction, in order to confirm the optical/UV results. Among them is the present study using the molecular gas mass measured by the CO emission line. In Table \ref{tab_studies}, we present a list of GRB host galaxies whose SFRs are measured in the optical/UV band and at least one other extinction-free method. As can be seen, these results do not necessarily coincide with each other.

Figure \ref{fig_SFR} shows how the SFR of GRB host galaxies in Table \ref{tab_studies} measured by extinction-free methods appear when compared to their SFRs measured in the optical/UV band. As can be seen, most of the results (positive detections as well as upperlimits) of extiction-free methods demonstrate SFRs that are greater than the optical results by one to a few orders of magnitude. The ``positive detections'' that are especially above the one-to-one correspondence line are the results of submillimeter and radio observations by \citet{2003ApJ...588...99B}, where some of these call into question by the much lower ``upperlimits'' set by the non-detection in mid-IR observations \citep{2006ApJ...642..636L}. In order to confirm whether GRB host galaxies are actually sub-luminous dwarf galaxies as they appear in the optical view, or dusty  active star forming galaxies as they seem to appear in the submillimeter band, measurements with sensitivity to star forming activity as high as the optical/UV methods must also be performed in other wavelengths. The CO measurement has the potential of becoming one of such methods, especially by virtue of future instruments with higher sensitivity, such as the Atracama Large Millimeter-and-submillimeter Array (ALMA).

\begin{deluxetable*}{cccccccc}


\tabletypesize{\tiny}


\tablecaption{Star Formation Rate of GRB Host Galaxies}

\tablenum{5}

\tablehead{\colhead{Source} & \colhead{Unobscured SFR} & \colhead{Tracer} & \colhead{Ref.} & \colhead{Optical SFR} & \colhead{Tracer} & \colhead{Extinction} & \colhead{Ref.} \\ 
\colhead{} & \colhead{($M_{\sun}\ \mathrm{yr^{-1}}$)} & \colhead{} & \colhead{} & \colhead{($M_{\sun}\ \mathrm{yr^{-1}}$)} & \colhead{} & \colhead{corrected} & \colhead{}  \\ 
\colhead{(1)} & \colhead{(2)} & \colhead{(3)} & \colhead{(4)} & \colhead{(5)} & \colhead{(6)} & \colhead{(7)} & \colhead{(8)} } 

\startdata
GRB 970228 & $<$335 & Submm & 1 & $\sim$0.76 & [OII] & no & 2 \\
GRB 970508 & $<$380 & Submm & 1 & $\sim$10 & $\mathrm{[OII]}$ & yes & 4 \\
 & $<$17 & IR & 3 &  &  &  &  \\
GRB 970828 & $80 \pm 60$ & Radio & 1 & $\sim$1.1  & $\mathrm{[OII]}$ & no & 5 \\
 & $24^{\ +\ 43}_{\ -\ 14}$ & IR & 3 &  &  &  &  \\
GRB 971214 & $120\pm 275$ & Submm & 1 & $\sim$5.2 & UV & no & 6 \\
GRB 980425 & 0.4 & IR & 3 & $\sim$0.35 & $\mathrm{H\alpha}$ & no & 8 \\
 & $<$2.8 $\pm 1.9$ & X & 7 &  &  &  &  \\
GRB 980613 & $380 \pm 200$ & Submm & 1 & $\sim$5.6 & $\mathrm{[OII]}$ & no & 9 \\
 & $50 \pm 140$ & Radio & 1 &  &  &  &  \\
 & $87^{\ +\ 156}_{\ -\ 52}$ & IR & 3 &  &  &  &  \\
GRB 980703 & $<$380 & Submm & 1 & $\sim$14 & $\mathrm{H\beta}$ & yes & 4 \\
 & $180 \pm 25$ & Radio & 1 &  &  &  &  \\
 & $<$24 & IR & 3 &  &  &  &  \\
GRB 981226 & $150 \pm 85$ & Radio & 1 & $1.2 \pm 0.3$ & UV & no & 10 \\
GRB 990123 & $<$140 & IR & 3 & $\sim$3.6-4.6 & UV & no & 11 \\
GRB 990506 & $<$170 & IR & 3 & $\sim$12.6 & $\mathrm{[OII]}$ & no & 12 \\
GRB 990705 & $190 \pm 165$ & Radio & 1 & $\sim$5--8 & UV & no & 13 \\
 & $32^{\ +\ 37}_{\ -\ 11}$ & IR & 3 &  &  &  &  \\
GRB 991208 & $<$370 & Submm & 1 & $ \sim$156 -- 249 & $\mathrm{H\beta}$ & yes & 4 \\
 & $70 \pm 30$ & Radio & 1 &  &  &  &  \\
GRB 000210 & $560 \pm 165$ & Submm & 1 & $\sim$3 & $\mathrm{[OII]}$ & yes & 14, 15 \\
 & $90 \pm 45$ & Radio & 1 &  &  &  &  \\
GRB 000418 & $690 \pm 195$ & Submm & 1 & $\sim$15.4 & $\mathrm{[OII]}$ & yes & 16 \\
 & $330 \pm 75$ & Radio & 1 &  &  &  &  \\
GRB 000911 & $495 \pm 195$ & Submm & 1 & $\sim$2.7 & UV & yes & 17 \\
 & $85 \pm 70$ & Radio & 1 &  &  &  &  \\
GRB 000926 & $820 \pm 340$ & Radio & 1 & $\sim$24 & UV & yes & 18 \\
GRB 010222 & $610 \pm 100$ & Submm & 1 & $1.5$ & & no & 1 \\
 & $300 \pm 115$ & Radio & 1 &  &  &  &  \\
 & $<$130 & IR & 3 &  &  &  &  \\
GRB 030329 & $<$4.0 -- 38 & CO & 19 & $\sim$0.6 & $\mathrm{[OII]}$ & yes & 20 \\
GRB 031203 & $<$150 $\pm 110$ & X-ray & 7 & $>$11 & $\mathrm{H\alpha}$ & yes & 21 \\
\enddata


\tablecomments{
Col. (1): Source name; 
col. (2): SFR derived from tracers free from dust extinction; 
col. (3): SFR tracer of col. (2); 
col. (4): References to col. (2); 
col. (5): SFR derived from optical tracers; 
col. (6): SFR tracer of col. (5). [OII] is the 3727 $\mathrm{\AA}$ doublet; 
col. (7): Whether the SFR in col. (5) is corrected for both local and Galactic dust extinction or not.
col. (8): References to col. (5).
\label{tab_studies}
}

\tablerefs{
1)\citet{2003ApJ...588...99B}; 
2)\citet{2001ApJ...554..678B}; 
3)\citet{2006ApJ...642..636L}; 
4)\citet{2001A&A...372..438S}; 
5)\citet{2001ApJ...562..654D}; 
6)\citet{1998Natur.393...35K}; 
7)\citet{2004A&A...425L..33W}; 
8)\citet{2005NewA...11..103S}; 
9)\citet{2003ApJ...591L..13D}; 
10)\citet{2005ApJ...631L..29C}; 
11)\citet{1999ApJ...518L...1B}; 
12)\citet{2003AJ....125..999B}; 
13)\citet{2002ApJ...581L..81L};  
14)\citet{2002ApJ...577..680P}; 
15)\citet{2003A&A...400..127G}
16)\citet{2003A&A...409..123G};  
17)\citet{2005A&A...438..841M}; 
18)\citet{2001A&A...373..796F}; 
19)this work.
20)\citet{2005A&A...444..711G};  
21)\citet{2004ApJ...611..200P}.}
\end{deluxetable*}

\section{CONCLUSION}

No CO ($J$ = 1--0) emission was detected from the GRB 030329 host galaxy.
The 3$\sigma$ upperlimit on the CO ($J$ = 1--0) line luminosity of the host galaxy is $3\cdot \delta L\arcmin_{\mathrm{CO}} = 6.9\times 10^8 \mathrm{\ K\  km\  s^{-1}\  pc^2} $. 
The lowerlimit of the host galaxy's metallicity is estimated to be $12+\log(\mathrm{O/H})\sim 7.9$ ,
which yields a CO line luminosity to $\mathrm{H_2}$ conversion factor of $\alpha_{\mathrm{CO}} = 40 \ M_{\mathrm{\sun}}\ (\mathrm{K\ km\ s^{-1}\ pc^2})^{-1}$.
Assuming this $\alpha_{\mathrm{CO}}$ factor, the 3$\sigma$ upperlimit on the molecular gas mass of the host galaxy is $3\cdot \delta M_{\mathrm{H_2}} = 2.8 \times 10^{10}\  M_{\mathrm{\sun}}$. 
Using the Schmidt law, the 3$\sigma$ upperlimit on the total star formation rate of the host galaxy is estimated to be $3\cdot \delta \mathrm{SFR} = 38 \ M_{\mathrm{\sun}}\mathrm{\ yr^{-1}}$.
Though the host galaxy of GRB 030329 is most likely a compact dwarf galaxy as seen in the optical band, multi-wavelength observations of GRB host galaxies are still essential in estimating their SFR accurately in order to reveal their star forming nature. The CO line observations will play an important role as an independent SFR estimator, free of dust extinction and ambiguity of dust temperature.

\acknowledgments
We would like to acknowledge the referee for his invaluable comments.
We wish to thank the NRO staff for operating the telescopes and for their continuous efforts to improve the performance of the instruments. NRO is a branch of the National Astronomical Observatory, National Institute of Natural Sciences, Japan. 
This study was financially supported 
by the MEXT Grant-in-Aid for Scientific Research 
on Priority Areas No.\ 15071202.
A. E., T. O., and K. M. were financially supported by the Japan Society for the Promotion of Science (JSPS) for Young Scientists.

{\it Facilities:} \facility{NoMA ()}

\clearpage




\end{document}